\documentclass[pra,amssymb,amsmath,showkeys]{revtex4}
\usepackage{graphicx}
\usepackage{color}

\usepackage{ulem}

\newcommand{\bc}{\begin{center}}
\newcommand{\ec}{\end{center}}

\begin{document}

\title{Uncomputability and free will} 

\author{Chetan
  S.  Mandayam Nayakar\footnote{aka M.  N. Chetan Srinivas}}
\email{mn_chetan@yahoo.com}, 
  % Srikrishna,     
\author{R.      Srikanth}\email{srik@poornaprajna.org}
\affiliation{Poornaprajna   Institute    of   Scientific   Research,
    Sadashivnagar,   Bangalore,  India}
\affiliation{Raman  Research
    Institute, Sadashivnagar, Bangalore, India}

\begin{abstract}
The basic  problem posed by free  will (FW) for physics  appears to be
not the  \textit{physical} one  of whether it  is compatible  with the
laws of physics,  but the \textit{logical} one of  how to consistently
define  it, since  it incorporates  the contrary  notions  of freedom,
which  suggests  indeterminism, as  well  as  control, which  bespeaks
determinism.   We argue  that it  must be  a fundamentally  new causal
primitive,   in  addition  to   determinism  and   indeterminism.   In
particular, we identify FW in  a physical theory with dynamics that is
uncomputable, and hence effectively indeterministic within the theory.
On the other hand, it would be deterministic in a higher order theory.
An important consequence for  artificial intelligence (AI) is that the
FW aspect  of cognitive systems may be  fundamentally unsimulable.  An
implication  for   neuroscience  is  that   FW  will  in   general  be
experimentally   undemonstrable.    Apparently,   it   can   only   be
subjectively experienced.
\end{abstract}
% \tableofcontents
\keywords{Uncomputability, undecidability, free will, Turing machines}
\maketitle

\section{Introduction\label{sec:intro}}
We  informally understand free  will (FW)  as the  power to  choose an
alternative from  many others.  However  for over two  thousand years,
philosophers, scientists  and theologians have  debated on what  FW is
and whether  it exists \cite{tim,fw,linda,ch}. A  basic difficulty has
been that  FW incorporates two opposing notions:  freedom and control.
On the one hand,  freedom suggests unpredictability and indeterminism,
whereas on  the other hand,  control suggests intent  and determinism.
From this perspective, FW is an oxymoron.

Two  broad philosophical positions  on FW  are \textit{compatibilism},
according   to  which   determinism   is  compatible   with  FW,   and
\textit{incompatibilism},   according    to   which   the    two   are
incompatible. According to a compatibilist, a person may choose freely
and yet an omniscient being  may possess foreknowledge of that choice.
An  incompatibilist  may  reject  FW,  in  which case  he  is  a  hard
determinist, or reject determinism, in which case he is a libertarian.

% theological 

FW has figured in some recent discussion in the foundations of quantum
mechics   (QM),  where   it   is   treated  simply   as   a  form   of
unpredictability,  or  independence from  all  past information  (with
`past'   defined  in   an  appropriately   relativistically  invariant
way).  Some works  in physics  that have  taken a  closer look  at the
existence   or  non-existence  of   FW  include   Refs.   \cite{stapp,
  suControl, hoo07,  vedral, gisin3440, sabine,  svet}.  The important
contribution of quantum  mechanics to this debate is  in introducing a
concrete instance of fundamental indeterminism via the $|\psi|^2$ Born
rule.

In this article, we hope  to characterize libertarian FW.  The article
is structured  as follows. Section \ref{sec:wfwp}  studies the problem
of  accommodating  FW  in  physical  laws.   Section  (\ref{sec:sfwp})
examines  the  logical  problem  of reconciling  unpredictability  and
control.  Section \ref{sec:uf}  develops a model of FW,  in which this
reconciliation is achieved by modelling  FW as a new causal primitive,
or fundamental  kind of causation,  which is not computable  and hence
deterministic within  the theory.  The  model is applied  to practical
situations to  eludidate its  resolution of the  paradox posed  by FW.
Finally,  in Section \ref{sec:ain},  we look  at implications  for AI,
brain science and neuroscience.

\section{Free Will and physical laws\label{sec:wfwp}}

If libertarian FW exists, then  determinism is ruled out as universal.
Is  the lot  of indeterminism  any  better? Intuitively,  it seems  so
because the indeterminacy gives gives `elbow room' for FW to act.  But
neither  does indeterminism  (as  governed by  some fixed  probability
rule, $P$) leave enough room for FW to act.

Consider the sample mean $X_n$ over $n$ trials
$$\lim_{n                      \rightarrow                     \infty}
\textrm{Pr}\left(\left|\overline{X}_n-\mu\right|>\epsilon\right)=0,
$$ by  the Weak  Law of Large  Numbers. This  implies that there  is a
`probability   pressure'  \textit{not}  to   choose  \textit{atypical}
sequences, which  would cause deviations from the  sample mean.  Thus,
there is a kind of long-run determinism, and hence a restriction on FW
as  we intuitively  understand it.   

Therefore, if  we accept libertarian  FW, the free-willed  choice will
potentially intefere with the underlying physical dynamics ${\cal D}$.
This interference will  take the form of (a)  overriding causality, if
${\cal  D}$  is deterministic;  or  (b)  causing  deviations from  the
relevant  probability  rule $P$,  if  ${\cal  D}$ is  indeterministic.

Clearly, it is immaterial  whether the underlying physics is classical
or quantum.  FW  itself can't be part of the  dynamics ${\cal D}$, for
in that case it could not produce the required deviations.  Therefore,
we require a Cartesian dualism with a physical and an `extra-physical'
component making up a free-willed agent.  Physical dynamics ${\cal D}$
governs the former while FW comes from the latter.  It is important to
stress  that  this extraphysical  agency  must  be  something that  is
\textit{qualitatively} different.

\section{The Fundamental problem of Free Will \label{sec:sfwp}}

The basic  problem of FW  is how to  reconcile the notion  of freedom,
which  suggests unpredictability  and indeterminism,  with  the notion
control,   which  suggests  the   opposite.  The   following  argument
highlights  this  difficulty:   (I)  Given  a  set  $X   =  \{x\}$  of
possibilities,  suppose the  agent $A$  chooses $y  = \mathcal{A}(X)$,
where $\mathcal{A}$  is an  algorithm representing $A$'s  mechanism of
choosing;  (II)   The  principle  of  excluded   middle  implies  that
$\mathcal{A}$ is deterministic or it is not; (III) In the former case,
there is no genuine situation involving choice, and hence no FW.  (IV)
In the  latter case, there is  only randomness, and  hence no control,
and no FW.

We thus find again that the agent's choice is deterministic or random,
with no apparent  room for (libertarian) FW. To  us, this suggests the
following model as a way out, in which the basic insight is that FW is
not about  unpredictability, but  the power to  make a  choice against
certain odds or compulsions.

Given the state $\rho$ of an agent, he is subject to two influences:
\begin{description}
\item{\bf  Contraints:} coming from  Nature, in  the form  of desires,
  instinctive drives  and emotional tendencies. This  corresponds to a
  computable function, described by  Turing machine (TM) $C$ acting on
  an encoding of state $\rho$ (informally written $C(\rho)$).
\item{\bf  Guidance:} coming  from a  rational capacity  to  model the
  world,  and to  understand  the (ethical,  social, financial,  etc.)
  implications  of  each choice.   It  is  described  by a  computable
  function, computed by TM $G$, and has the action $G(\rho)$.
\end{description}
Here $\rho$ may be thought of as an integer, say the G\"odel or Turing
number, that  encodes the physical  state.  In the simplest  case, the
dilemma  facing the  agent is  whether to  go with  $C(\rho)$  or with
$G(\rho)$.

Suppose  there is  a situation  containing  alternative possibilities,
labelled $x \in X$. If there is no such thing as FW, then the combined
action of  $C$ and $G$  fully determine the  outcome $x$. This  may be
expressed by saying that
\begin{equation}
\rho^\prime = \phi\left(C(\rho),G(\rho)\right),
\label{eq:fc}
\end{equation}
where $\phi$ is a computable  function, that may be linear, nonlinear,
deterministic or  probabilistic.  In this  case, the output  state can
computed,   and  the  dynamics   represented  by   Eq.   (\ref{eq:fc})
understood.  Such  a model is  suitable for a compatibilistic  or hard
deterministic  world view. One  can propose  a more  involved function
that Eq. (\ref{eq:fc}), such as:
\begin{equation}
\rho^\prime = \chi(C,U)(\rho),
\label{eq:fq}
\end{equation}
but the former suffices for our purpose.

If  we  accept the  libertarian  position,  then  a third  element  is
required, which expresses  the idea of empowering the  agent to make a
choice  under  the  (possibly  opposing) influences  of  $C(\rho)$  or
$G(\rho)$.  This is the faculty of volition, or:
\begin{description}
\item{\bf Freedom of Will.}  The freedom to orient or align the choice
  in  line  with  the   Understanding  $G(\rho)$,  by  overcoming,  if
  necessary, the  Constraint $C(\rho)$.  We represent it  by a process
  $\mathcal{A}$.
\end{description}
For FW in  this sense to be tenable,  the process $\mathcal{A}$ itself
should  be extra-physical, as  noted earlier.   In particular,  in Eq.
(\ref{eq:fc}), function  $\phi$ should  be uncomputable in  the Turing
model.   This  can  be  shown   in  analogy  with  the  proof  of  the
uncomputability of the halting problem for TMs \cite{turing}.

\section{Uncomputability and FW \label{sec:uf}}

Suppose prediction $Y$ is made  about agent $S$.  If $S$ believes $Y$,
$S$ can falsify  it by deliberately acting contrary to  $Y$. It may be
supposed  that a  more  detailed  algorithm to  arrive  at an  updated
prediction  $Y^\prime$  should be  able  to  take  into account  $S$'s
reaction.  Yet  $S$ may simply choose to  falsify \textit{that}.  Only
if $S$  disbelieves $Y$ will  $Y$ hold with certainty.   However, this
would make $S$ a sort of inconsistent agent, for disbelieving a truth.
Thus,  if $S$  is consistent,  then $S$  must believe  $Y$.   To avert
inconsistency, we conclude that in  general there is no algorithm that
can predict how an agent will behave.  The similarity of this argument
to  G\"odel's  celebrated  incompleteness theorem  \cite{god}  becomes
apparent when we replace the  notion of provability there with that of
belief.   An agent's  behavior  may appear  random  because it  cannot
always be predicted, but is  not (in this model) ontologically random.
It is  this ability  to act in  an unpredictable, yet  non-random, way
that we deem FW.

The  similarity  to  the  halting   problem  for  TMs  is  also  clear
\cite{turing}. Consider the choice  problem $H_A$, mentioned above, of
predicting the choice of an agent. The undecidability of $H_A$ follows
from the  reduction of Turing's Halting  problem $H$ to  $H_A$.  If we
had  a  decider $\mathcal{A}$  for  $H_A$  (one  that can  answer  all
instances of  $H_A$ in finite time), then  one, denoted $\mathcal{H}$,
for $H$ can  be constructed as follows: $\mathcal{H}$  constructs a TM
$N$ that  outputs ``0'' if  a TM $M$  halts on input $w$,  and outputs
``1''   otherwise.   To   decide,  $\mathcal{H}$   can   now  evaluate
$\mathcal{A}(N)$ and determine  that $M$ halts (does not  halt) on $w$
if the  output is  $0$ $(1)$.  A  non-reductive proof is  also readily
obtained via the usual diagonal argument.

Suppose  an algorithm  $\mathcal{A}$  exists that,  knowing the  state
$\rho$ of  an agent $S$, can  predict in finite time  whether $S$ will
choose ``0'' or ``1'' on input $i$. That is,
\begin{equation}
\mathcal{A}(S; i) =
\left\{
\begin{array}{cc}
0 \Longleftrightarrow S(i) = 0 \\
1  \Longleftrightarrow S(i) = 1
\end{array} \right.
\label{eq:2}
\end{equation}
Then one can construct the  program $\mathcal{R}$, which represents an agent
who reacts to a prediction about herself by falsifying it, as follows:
\begin{equation}
\mathcal{R}(p) = \left\{
\begin{array}{cc}
1 \Longleftrightarrow \mathcal{A}(p;p) = 0 \\
0  \Longleftrightarrow \mathcal{A}(p;p) = 1
\end{array} \right.
\label{eq:3}
\end{equation}
Applying $\mathcal{A}$ to $\mathcal{R}$ we have from Eqs. (\ref{eq:2})
and (\ref{eq:3}):
\begin{equation}
\mathcal{A}(\mathcal{R}; p) = \left\{
\begin{array}{cc}
1 \Longleftrightarrow \mathcal{A}(p;p) = 0 \\
0  \Longleftrightarrow \mathcal{A}(p;p) = 1
\end{array} \right.
\label{eq:4}
\end{equation}
Here we have assumed that  $\mathcal{R}, p$, etc.  as arguments, stand
for integer  labels for the corresponding  quantity.  Eq. (\ref{eq:4})
leads  to a  contradiction when  we set  $p =  \mathcal{R}$,  i.e., we
choose $p$ to be the  Turing number of $\mathcal{R}$. We conclude that
to avoid it, $\mathcal{A}(\mathcal{R}; \mathcal{R})$ loops infinitely.

In particular,  the performance of  $\mathcal{A}$ fails in  cases like
$\mathcal{R}(\mathcal{R})$,  in   that  its  truth   value  cannot  be
consistently computed by the agent.  We can avert the inconsistency by
stipulating that  $\mathcal{A}(\mathcal{R}; \mathcal{R})$ never halts,
so  that $\mathcal{R}(\mathcal{R})$  is  not decidable,  which is  the
G\"odel  sentence \cite{god}  for the  formal system  encompassing the
agent, making it syntactically incomplete. G\"odel undecidability thus
is  implied by Turing  uncomputability \cite{chaitin}.  Detangling the
argument of Eq.   (4), we find that a  general predictive algorithm is
logically impossible because it would help create an algorithm that is
so powerful that,  knowing its own future, it  could act contradictory
to its own prediction.  Consistency dictates that arbitrarily powerful
self-predicting   programs   are    impossible.    The   question   of
self-reference in  a FW-endowed entity was first  considered by MacKay
\cite{mackay,  barrow}, though  he  denied its  connection to  G\"odel
incompletess or Turing non-computability.

If  we assume  that all  properties of  an agent  that  are physically
directly accessible  are computable, then  the component of  the agent
(which  we may  call ``mind'')  responsible for  process $\mathcal{A}$
must be something that is  physically not directly accessible, and can
be considered extra-physical in  that sense. Something about the brain
allows the  mind to interface with  matter in a way  to allow FW.

FW  as   defined  above  is  the  new   \textit{causal  primitive}  or
\textit{principle   of   causation},   apart  from   \textit{physical}
determinism  and  indeterminism.  Because  the  full  dynamics is  not
accessible, the  map which represents  the act of choosing  an element
from  a  set, the  physically  accessible  free-willed behavior,  will
appear  seem  probabilistic.  Causality  is  no  longer  closed  under
physics.

A simple way to model this is  as follows. Let $|X| = d$. We represent
$\rho$ by a stochastic column  vector $\vec{\rho}$, and $C$ and $G$ be
right stochastic  matrices $\vec{C}$  and $\vec{G}$. The  evolution of
the state is given by:
\begin{equation}
\vec{\rho^\prime}           =           \left(\alpha\hat{G}          +
(1-\alpha)\hat{C}\right)\vec{\rho} \equiv \hat{U}\vec{\rho},
\label{eq:evo}
\end{equation}
where $0 \le \alpha \le 1$. The stochastic matrix $\hat{U}$ represents
the  free-willed  action and  encompasses  the  combined influence  of
guidance  and constraints of  nature.  The  number $\alpha$  and hence
$\hat{U}$  are  in general  uncomputable  \cite{chaitin},  but can  in
principle be estimated experimentally.

The larger the `FW parameter' $\alpha$, the more is the agent's choice
geared according to  guidance $G$, and less the  compulsion of nature.
As a  simple example, given  a dichotomic choice space  $\{0,1\}$, let
$G$  recommend choosing  0,  while  $C$ recommend  1.   For any  state
$\vec{\rho} = (p,1-p)^{T}$, we find
\begin{equation}
\vec{\rho^\prime} = 
\left( \begin{array}{c} \alpha \\ 1-\alpha
\end{array}\right)
= \left[ \alpha\left( \begin{array}{cc}
1 & 1 \\ 0 & 0
\end{array} \right)
+
(1-\alpha)\left( \begin{array}{cc}
0 & 0 \\ 1 & 1
\end{array} \right) \right] \vec{\rho}
\label{eq:fw}
\end{equation}

\section{Applications \label{appli}}

We will now see how  this model resolves the logical paradox mentioned
above.  We  define a \textit{Saint}  as an agent  whose Understanding,
and hence guidance $G$, is  ethical and whose FW parameter $\alpha$ is
near maximal.   Consider the proposition: ``Presented  with the choice
between  the good and  the evil,  the Saint  {\it freely}  chooses the
good.''   In Eq.   (\ref{eq:fw}),  setting  0 as  ``the  good'', 1  as
``evil'', and  $\alpha=1$, we find $\vec{\rho^prime}  = (1,0)^T$.  The
Saint, by dint  of high FW, can if  necessary override any Constraints
imposed by  his human  nature, to always  choose the good.   Thus, the
deterministic  behavior  and  predictability  of the  Saint  does  not
preclude  his  free  will,  contrary to  the  \textit{incompatibilist}
implication (III)  above.  On the other end,  predictable behavior can
arise also  because of  low degree  of FW, which  obtains when  we set
$\alpha=0$ in Eq. (\ref{eq:fw}).

Suppose that an agent's FW is not maximal.  Then his choice fluctuates
randomly between choosing according to  the dictates of his nature and
the recommendation  of his Understanding.  This is  illustrated in the
following proposition, whose intuitvely apparent truth is validated by
our definition of FW: ``Presented with the choice between the good and
the evil, the Conscientious Criminal vacillates'' This criminal, being
conscientious,  has a  clear Understanding  of the  virtue  of ethical
behavior, but,  owing to lack  of sufficiently high FW,  cannot always
overcome   the   compulsion   of   his  criminal   nature.    In   Eq.
(\ref{eq:fw}),   setting  $\alpha=0.5$,  $\vec{\rho^\prime}   =  (0.5,
0.5)^{T}$.  His choice is random,  being good sometimes and being evil
at other times.  Thus the randomness of his choice does not imply lack
of free will.   It only implies low free will.   This explains why the
\textit{pessimist} \cite{fw} implication (IV) fails.

\section{Implications for AI and neuroscience \label{sec:ain}}

Our analysis suggests that if we accept the proposition of libertarian
FW,  then something  extra-physical or  mind-like, and  in particular,
uncomputable, happens in  the act of volition, at  the instant where a
new intent is  generated in the brain. Any  AI attempt to algorithmize
consciousness must then  fail because true FW cannot  be simulated, at
least  in machines that  implement first  order logic.   FW is  then a
facet  of  conscious  beings  that   seems  to  set  them  apart  from
automatons.

Darwinists may  argue that  FW is a  strategy necessary to  prevent an
organism or  specie from settling down  in local minima  in the energy
landscape appropriate to biological thermodynamics, and thus to ensure
long-term  self-preservation. A  patient who  submits to  a short-term
painful treatment for good health in  the long run, or an altruist who
starves  to feed  the needy,  are,  in this  view, strategies  reached
through FW that could otherwise  have been missed.  In this view (that
is  freely  held  by  one   of  the  authors--  CMN!),  FW  originated
evolutionarily to enable deviation from the local gradient and thereby
do  better  than  locally  optimize  in  an  organism's  struggle  for
self-preservation. For example, in  a mountaineous terrain, the higher
the oxygen  concentration the better,  and the lower  the geographical
altitude, the  higher the oxygen concentration.   Organisms without FW
would follow the  local gradient and settle down  in the local minimum
valley.  The intervention  of FW will let an organism  to climb over a
mountain peak in order to reach  a deeper valley.  The evolution of FW
also brought along ``by-products'' like altruism or self-harm.

The subject of FW rightly receives much attention in brain science and
neuroscience  \cite{libet,  trevena,  rosk}.   For  neuroscience,  the
implication  of our  findings  is that  any  attempt to  trace back  a
particular action of an organism to  its first cause in a neuron (such
as pyramidal neurons identified as the seat of origin of mouse whisker
twitchings   \cite{mouse}),    or   in   a    sub-neuronal   structure
\cite{penham}, is bound to fail.  It may be preserved to a great depth
as one traces back down the  signal pathway, but there will be a final
`flicker'  that is irreducible  to other  observable causes,  and thus
visibly  random, and  yet quite  rational when  the bigger  picture is
taken into account.

But even  to get there,  an experimental test would  require isolating
systems of  interest to rule  out extraneous stimuli.   Otherwise, one
could  not rule  out a  conventional, physical  explanation,  based on
classical indeterminism \cite{brembs} or even quantum indeterminism as
being responsible for observed fluctuations in behavior.

In this  sense, FW will fundamentally  be experimentally inaccessible,
and thus  not scientifically  falsifiable.  One can  only subjectively
experience it, and also rule  out an action as being free-willed (when
it is manifestly explained by  visible causes), but never point out FW
directly. One can at best look for circumstantial evidence.

Our resolution of the  problem of FW is to identify it  with a kind of
uncomputable  determinism. It  will be  computable if  one  can access
oracle machines, which are TMs equipped  with a black box that is able
to answer  in one  step any  instance of the  halting problem.   But a
halting  problem, or  equivalently the  FW problem,  exists  for these
oracle  machines, when  they  are applied  to  machines equivalent  to
themselves.   One can  thus define  a second-order  FW.  Extending the
notion  of  oracle  machines,  one obtains  the  arithmetic  hierarchy
\cite{ah},  where each  higher  stage brings  a  more powerful  oracle
machine and an even harder halting problem.  It is thus an interesting
question whether absolute FW exists!


\begin{thebibliography}{99}
\bibitem{suControl} A. Suarez. Quantum randomness can be controlled by
  free   will–-  a  consequence   of  the   before-before  experiment.
  arxiv:0804.0871.
\end{thebibliography}
\end{document}